\documentclass[aps,prc,twocolumn,showpacs,superscriptaddress]{revtex4-1}
\usepackage{graphicx}
\usepackage{dcolumn}
\usepackage{bm}

\begin{document}

\title{Systematics  of  one-quasiparticle configurations  in 
       neutron-rich Sr, Zr, and Mo odd isotopes with the Gogny 
       energy density functional.}

\author{R. Rodriguez-Guzman}
\author{P. Sarriguren}
\affiliation{Instituto de Estructura de la Materia, CSIC, 
Serrano 123, E-28006 Madrid, Spain}

\author{L.M. Robledo}
\email{luis.robledo@uam.es}
\homepage{http://gamma.ft.uam.es/~robledo}

\affiliation{Departamento  de F\'{\i}sica Te\'orica, 
Universidad Aut\'onoma de Madrid, 28049-Madrid, Spain}

\date{\today}

\begin{abstract}

The systematics of one-quasiparticle configurations in neutron-rich 
Sr, Zr, and Mo odd isotopes is studied within the Hartree-Fock-Bogoliubov 
plus Equal Filling Approximation method preserving both axial and 
time reversal symmetries. Calculations based on the Gogny energy density 
functional with both the standard D1S parametrization and the new D1M 
incarnation of this functional are included in our analysis. The nuclear 
deformation and shape coexistence inherent to this mass region are shown 
to play a relevant role in the understanding of the spectroscopic features
of the ground and low-lying one-quasineutron states.

\end{abstract}


\maketitle

\section{Introduction}
\label{INTRO}

The theoretical description of odd-mass nuclei is a difficult task
in mean field models and, for that reason, they have been much
less studied than the corresponding even-even systems although 
they roughly constitute half of the existing nuclides.
The properties of odd nuclei are however relevant, for example,  to
understand odd-even effects related to pairing correlations, 
as well as to constrain nuclear
energy density functionals (EDFs) aiming to reach a reasonable
spectroscopic quality. Within this context, the energies, spins and parities
of the (neutron and/or proton) multi-quasiparticle 
excitations are basic pieces of information
to get insight into the underlying nuclear structure.
Thus, the interest in the properties of odd-$A$ nuclei and their
computational challenges have been lately revived
\cite{duguet01,bonneau,perez,hamamoto,schunck}.

One of the main technical difficulties in the description of odd nuclei
is that the (exact) blocking procedure \cite{Mang-PREPORT,rs}
requires the breaking of time-reversal invariance
making the calculations  troublesome. Moreover, since blocking does depend 
on the  choice of the alignment orientation, for each quasiparticle
excitation reorientation effects should also be considered \cite{Olbrat,schunck}
retaining in each case the solution with the lowest energy. Therefore, 
in order to reduce the computational cost, in practical applications it is 
convenient to preserve time-reversal as selfconsistent symmetry. 
This has been traditionally accomplished within the so called
Equal Filling Approximation (EFA), according to which the odd nucleon
is half into a given orbital and half into its time-reversed partner.
The microscopic justification of the EFA is based on standard ideas of 
quantum statistical mechanics and has been considered in
Ref. \cite{perez} . The predictions arising 
from  various treatments of blocking have been compared in  Ref. \cite{schunck}. On one hand, it 
has been shown that both  exact blocking  and  
EFA approach are  strictly equivalent when the time-odd fields of 
the EDF are neglected. On the other hand, it has also been concluded that the 
EFA is sufficiently 
precise for most practical applications. The EFA is precisely the approximation
employed in the present work to study the systematics of the 
ground and low-lying one-quasineutron states
in  odd-$A$ nuclei with proton (neutron) numbers $ 38 \le Z \le 42$ 
($ 47 \le N \le 67$).

The local systematics of one-quasiparticle excitations have already been studied
in various mass regions either from macroscopic-microscopic approaches
\cite{naza90,Ogle-1,WIOK-91,WIOK-94}  or from the EDF theory 
\cite{schunck,rutz,cwiok,satula-2008,rutz-EDF,Afa-1,Afa-2}. To the best 
of our knowledge, still the only global study  of odd-nuclei within the nuclear EDF framework 
has been reported  in Ref. \cite{bonneau} where 
Skyrme-HF+BCS predictions for  ground-state spins 
and parities have been compared with finite-range droplet model results.

As already mentioned, in this work we study neutron-rich Sr, Zr, and Mo 
odd isotopes with mass numbers $A \sim 100$. Nuclei in this region 
of the nuclear chart are being actively studied both theoretically 
\cite{skalski,xu02,oursplb,sarri} and experimentally 
\cite{buchinger,mach,urban, campbell,hua,goodin,charlwood,pereira}
because of their interest from the nuclear structure perspective, as
well as because of their significance in nuclear astrophysics. 
In particular, their masses and decay properties are an essential input to model the 
path, the isotopic abundances and the time scale of the r process in 
a reliable way \cite{cowan}. In addition, this region is also characterized 
by the strong competition between various shapes, giving rise to shape 
instabilities that lead to shape coexistence and 
sudden shape transitions \cite{wood92,oursplb}. Thus, it is interesting to explore 
the predictive power of the  Gogny-EDF \cite{gogny} 
when applied to describe the ground and low-lying one-quasineutron 
states in odd-$A$  nuclear systems 
belonging to this region. Although it is clear that one cannot expect 
{\it{a priori}} to reproduce in detail spectroscopic properties of the 
considered odd-$A$ nuclei from global EDFs (see, for
 example, \cite{Zala-other,Stoisov,Korta}) the question still remains 
about the ability of the Gogny-type EDFs  to describe, at least qualitatively, 
the main features of the observed trends  in the Sr, Zr and Mo isotopic chains. 

In addition to D1S \cite{d1s}, which is the most standard and thoroughly 
tested  parameter set (see, for example,  \cite{robledo_88,egido_92,egido-PRL,garrote,
mixed-dd-1,mixed-dd-2,mixed-dd-3,mixed-dd-4,rayner-PRL,
gogny-other-1,gogny-other-2,gogny-other-5,bertsch,peru,hilaire,delaroche} and references 
therein), in the 
present study we have also considered the most recent parametrization of 
the Gogny-EDF, namely the  D1M \cite{d1m} parametrization. The goal is twofold, first 
we want to verify the robustness of our predictions with respect to the particular
version of the EDF employed. If the results with two different EDFs (or two parametrizations
of one, as is our case) are similar, we can be rather confident about the independence
of the predictions with respect to  the details of the EDF.  
The second goal is to test the performance of D1M in the present context (spectroscopy of
odd-$A$ nuclei) as to add arguments to decide whether D1M \cite{d1m} can 
supersede the rather old D1S \cite{d1s} parametrization or 
not. The fitting protocol of D1M includes input from  a realistic Equation of State (EoS) both 
in symmetric and neutron nuclear matter. It also explores, in the symmetric nuclear matter
EoS  case,  the four possible spin-isospin channels and tries to reproduce 
the trends obtained in realistic Brueckner-Hartree-Fock calculations. As a
consequence, the D1M parametrization reproduces the correct sign for the isovector
splitting of the effective mass for neutron-rich matter at all possible asymmetries.
The consequence of such a sophisticated fitting protocol is an energy root mean square 
deviation for the experimentally known 2149 masses of only 0.8 MeV which is much better 
than the one of D1S.  The systematic comparison \cite{d1m} of the collective $2^+$ 
excitation energies obtained with D1S and D1M (only even-even nuclei were considered) 
suggests that the later essentially keeps the same nuclear structure predictive power 
of the former. Subsequent analysis centered on the structure of the intrinsic wave functions 
\cite{PT-D1M} has confirmed the previous findings. Therefore, it 
is timely and necessary to  further extend the comparison between D1S and D1M 
to odd nuclei. To this end, we analyze in the present paper the spectroscopic properties
of neutron-rich Sr, Zr and Mo odd isotopes in the  $A \sim 100$ mass region
within the Hartree-Fock-Bogoliubov (HFB) \cite{rs} plus  
EFA framework \cite{perez} (HFB-EFA).  The election of the considered  nuclei  is mainly driven by the 
intense experimental \cite{buchinger,mach,urban, campbell,hua,goodin,charlwood}
and theoretical \cite{skalski,xu02,oursplb,sarri}
efforts to better characterize the structural evolution of 
the ground and 
excited nuclear shapes in this region of the 
nuclear chart \cite{exp_spect}.

The paper is organized as follows. In Sec. \ref{THEORY}, we present a 
brief description of the theoretical formalism used in the present  work, 
i.e., the HFB-EFA framework. The results of our calculations for the considered 
nuclei are discussed in Sec. \ref{RESULTS} where  we pay attention to mean 
field potential energy curves (PECs), equilibrium deformations of the 
various competing shapes, single-particle energies, one-quasineutron
states and their spectroscopic evolution along the Sr, Zr and Mo isotopic 
chains. We also compare our results with the available 
experimental data. Finally, Sec. \ref{CONCLU}
is devoted to the concluding remarks and work perspectives.

%
%

\section{Theoretical framework}
\label{THEORY}

In the present study several Sr, Zr and Mo isotopes
with odd neutron number are studied. Nevertheless, the blocking procedure 
changing the number parity \cite{Mang-PREPORT,rs} of a given 
even-even HFB vacuum and providing us with the corresponding ground 
and low-lying one-quasiparticle states in odd-$A$ nuclei also requires to 
consider even-A nuclei in these isotopic chains, as discussed below.

In previous studies of even-even nuclei we have found advantageous
to use the so called gradient method \cite{PT-D1M,gradient,ours2} 
to obtain the solution of the HFB equations, leading to the (even number parity) 
vacuum  $| \Phi \rangle$. Within this method, the HFB equation is 
recast in terms of a minimization (variational) process of the mean field energy 
and the Thouless parameters defining the most general 
HFB wave functions \cite{rs} are used as variational parameters. For
a given point in the multidimensional variational space of the Thouless
parameters the procedure uses the direction of
the gradient in its search for the minimum. The advantage
of the gradient method over others, like 
the successive iteration method, relies on the way constraints are 
implemented, which allows a larger number of them to be treated at once.
As it is customary in calculations with the Gogny force, the 
kinetic energy of the center of mass motion has been subtracted 
from the Routhian to be minimized in order to ensure 
that the center of mass is kept at rest. The exchange 
Coulomb energy was considered in the Slater approximation and
we neglected the contribution
of the Coulomb interaction to the pairing field. Both axial and
time-reversal are selfconsistent symmetries in our calculations. 

The  quasiparticle creation and annihilation
operators $(\hat{\beta}_{k}^{\dagger}$ and $\hat{\beta}_{k})$ 
associated with a given (even-even)  HFB vacuum $| \Phi \rangle$ 
 have been expanded in an axially symmetric 
harmonic oscillator (HO) basis $(\hat{c}_{l}^{\dagger}, \hat{c}_{l})$

\begin{eqnarray} \label{Bogo-quasiparticles}
\beta^+_\mu &=& \sum_m U_{m\mu}c^+_m + V_{m\mu}c_m 
\nonumber\\
\beta_\mu &=& \sum_m U_{m\mu}^{*}c_m + V_{m\mu}^{*}c_m^+ .
\end{eqnarray}
containing enough number of shells (13 major shells) 
to grant convergence for all values considered 
of the mass quadrupole moment 
$Q_{20} = \frac{1}{2} \langle \Phi | 2z^{2} - x^{2} - y^{2}  | \Phi \rangle $
\cite{ours2} and for all the nuclei studied.
  
\begin{figure}[ht]
\centering
\includegraphics[width=80mm]{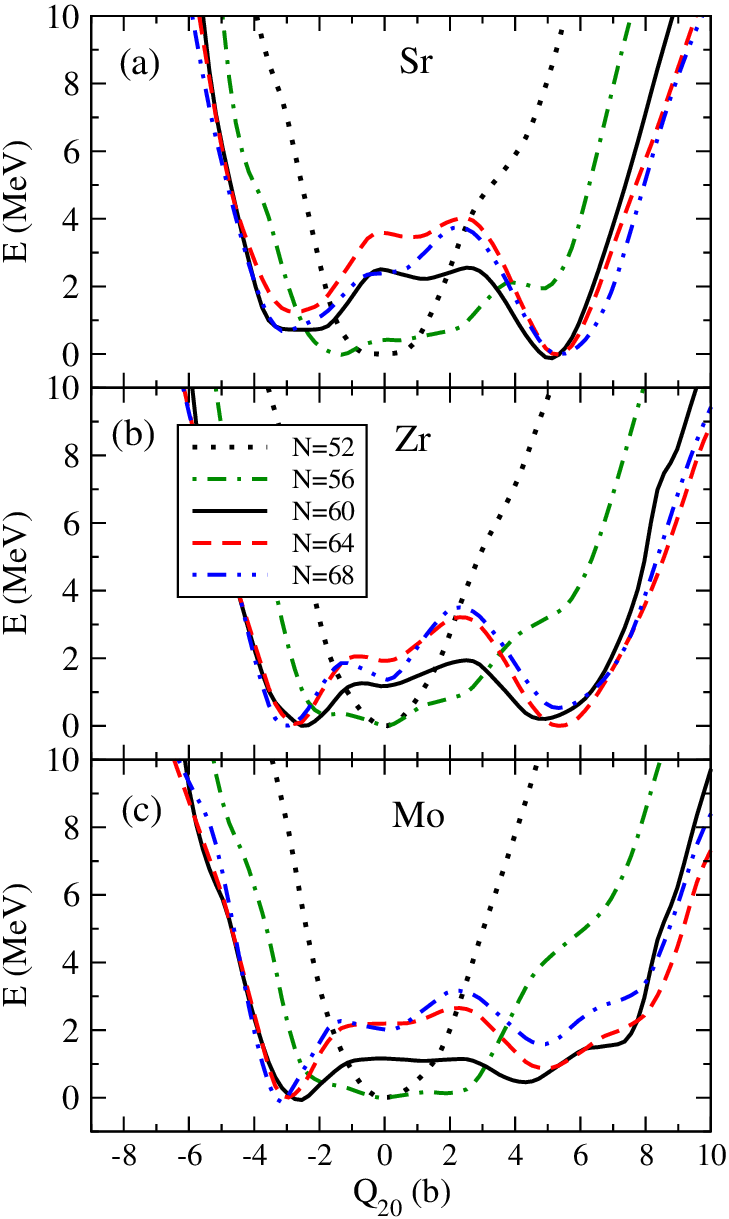}
\caption{(Color online) PECs for Sr (a), Zr (b), and Mo (c) isotopes with 
$N=$52, 56, 60, 64, and 68 obtained from HFB calculations with the 
Gogny-D1S energy density functional.}
\label{fig_pec}
\end{figure}

Constrained calculations have been performed to generate PECs for even-even
Sr, Zr and Mo nuclei. Such PECs are displayed in  
Fig. \ref{fig_pec} for a set of representative isotopes and 
the parametrization Gogny-D1S. One can also find
a systematic compilation of PECs obtained with Gogny-D1S in 
Ref. \cite{webpage}. The computation of such PECs is twofold:
first, they give us  initial hints on the evolution of the 
different competing shapes in the considered nuclei and, second,
they provide a whole set of prolate, spherical and oblate
(reference) even-even HFB states for the subsequent 
treatment of the neighboring odd-$N$ nuclei. 

In fact, once a reference (even-even) state with a given deformation is 
chosen, we use it to perform an additional (constrained) HFB calculation 
providing an unblocked fully-paired state corresponding to an odd average 
neutron number (false vacuum \cite{duguet01}) with the same deformation. 
Such  prolate, spherical and oblate false vacua are then used as input 
configurations in our subsequent blocking scheme (i.e., EFA). 

The HFB ground-state wave function $ |\Phi \rangle $ of an even-even nucleus 
is defined by the condition  of being the vacuum of the annihilation 
quasiparticle operators $\beta_\mu$ defined in Eq. (\ref{Bogo-quasiparticles}).
Regarding observables (mean values) the HFB ground state of an even-even 
system is specified \cite{rs} by the (hermitian) density matrix 
\begin{equation}
\rho_{ij}=\left( V^* V^T \right)_{ij}
\end{equation}
and the (antisymmetric) pairing tensor 
\begin{equation}
\kappa_{ij}=\left( V^* U^T \right)_{ij}
\end{equation}
where the $U$ and $V$ amplitudes are those of Eq (\ref{Bogo-quasiparticles})
defining the Bogoliubov transformation. Applying any unitary
transformation to the right of  both amplitudes $U$ and $V$ leaves
both $\rho$ and $\kappa$ (and therefore any mean value) unaltered
whereas the quasiparticle operators are affected by the representation
of the unitary transformation in operator space. This invariance is not
maintained in the odd-$A$ case, as we will discuss below, implying conceptual
changes in the treatment of an odd-$A$ system as compared to the simpler
even-even case.

On the other hand, the ground and low-lying one-quasiparticle 
states of odd-$A$ systems, like the ones considered in the present work, 
can be handled with blocked (odd number parity \cite{Mang-PREPORT,rs})
HFB wave functions
 \begin{equation} \label{one-quasi-state}
| \Psi_{\mu_B} \rangle  = \beta ^+_{\mu_B} | \Phi \rangle 
\end{equation}
where ${\mu_B}$ indicates the quasiparticle state to be blocked and 
stands for the indexes compatible with the symmetries of the odd-nuclei,
such as the angular momentum projection $K$ and parity in the case of 
axial symmetry. In practice, the blocked state (\ref{one-quasi-state})
can be  accomplished by exchanging the columns labeled by the 
index ${\mu_B}$ in the HFB amplitudes $U$ and $V$ [see, Eq. (\ref{Bogo-quasiparticles})]
with the same columns in the matrices 
$V^{*}$ and $U^{*}$, respectively  \cite{Mang-PREPORT,rs,exch-columns-Ring}.

The  density matrix  and the pairing tensor corresponding
to the one-quasiparticle state (\ref{one-quasi-state}), read
\begin{eqnarray} \label{dm-odd-exact}
\rho_{ij}^{(\mu_B)}&=&\langle \Psi_{\mu_B} | c^+_j c_i | \Psi_{\mu_B} \rangle 
= \langle \Phi | \beta_{\mu_B}  c^+_j c_i \beta_{\mu_B}^{\dagger}| \Phi \rangle
\nonumber \\
&=& \left( V^* V^T \right)_{ij} + \left( U^*_{j\mu_B}U_{i\mu_B}-V_{j\mu_B}V^*_{i\mu_B}
\right)
\end{eqnarray}
and
\begin{eqnarray} \label{pt-odd-exact}
\kappa_{ij}^{(\mu_B)}&=&\langle \Psi_{\mu_B} | c_j c_i | \Psi_{\mu_B} \rangle 
= \langle \Phi | \beta_{\mu_B}  c_j c_i \beta_{\mu_B}^{\dagger}| \Phi \rangle
\nonumber \\
&=&\left( V^* U^T \right)_{ij} + \left( U_{i\mu_B}V^*_{j\mu_B}-U_{j\mu_B}V^*_{i\mu_B}
\right) .
\end{eqnarray}
They violate time-reversal invariance and so do the
HF and pairing fields associated with them, making the numerical calculation much more 
difficult than in the even-even case. These matrices are no longer 
invariant under unitary transformations of the $U$ and $V$ 
amplitudes applied to their right. As a consequence,
that degree of freedom has to be taken into account in the solution of the
HFB equations. A manifestation of this effect is the 
reorientation effect  \cite{Olbrat,schunck} discussed recently.

\begin{figure}[ht]
\centering
\includegraphics[width=80mm]{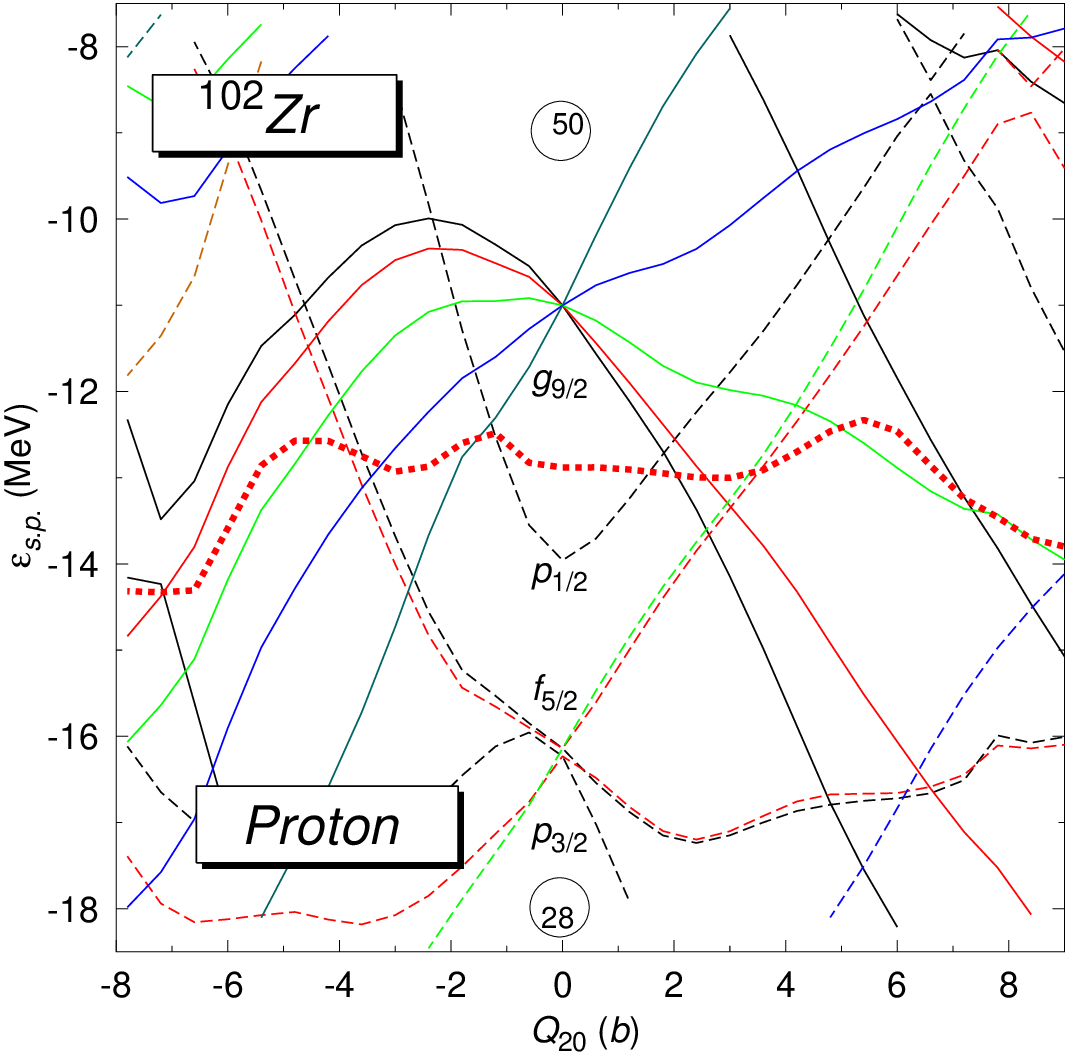}\par
\includegraphics[width=80mm]{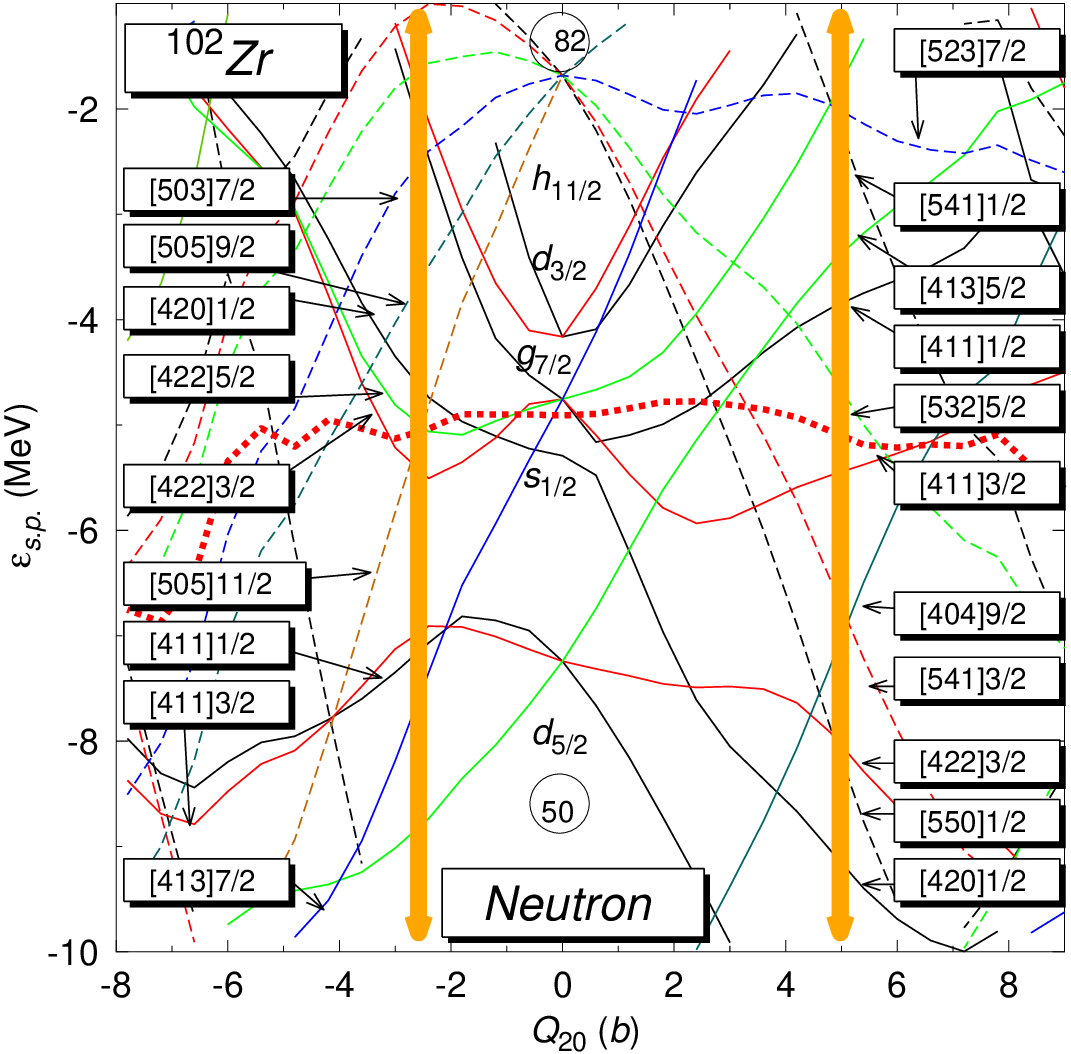}
\caption{(Color online) Single particle energies for protons and 
neutrons in $^ {102}$Zr as a function of the axial quadrupole moment 
$Q_{20}$. The Fermi level is depicted as a thick dashed red line. 
The results have been obtained with the Gogny-D1S EDF. 
Solid lines correspond to levels with positive parity whereas
dashed lines correspond to negative parity states.  
Asymptotic (Nilsson) quantum numbers $[N,n_z,\Lambda]K^\pi$ are shown in the
neutron case for  $Q_{20}$ values close to those where the minima of the PECs
are located (vertical arrows).
}
\label{fig_spe}
\end{figure}

A very useful approximation to (exact) blocking preserving 
time-reversal invariance is the EFA \cite{perez,schunck}, where 
one writes the density matrix 
\begin{eqnarray}
\rho_{ij}^{(EFA,\mu_B)} &=& \frac{1}{2} \left(
\langle \Phi | \beta_{\mu_B}  c_j^{\dagger} c_i \beta_{\mu_B}^{\dagger}| \Phi \rangle
+
\langle \Phi | \beta_{\overline{\mu}_B}  c_j^{\dagger} c_i \beta_{\overline{\mu}_B}^{\dagger}| \Phi \rangle
\right)
\nonumber\\
&=&
\left( V^* V^T \right)_{ij} +
\frac{1}{2}  \left( U_{i\mu_B}U^*_{j\mu_B}-V^*_{i\mu_B}V_{j\mu_B} \right)
\nonumber\\
&+&
\frac{1}{2}  \left( U_{i \overline{\mu}_B}U^*_{j \overline{\mu}_B}
-V^*_{i \overline{\mu}_B}V_{j \overline{\mu}_B} \right)
\end{eqnarray}
and  the pairing tensor 
\begin{eqnarray}
\kappa_{ij}^{(EFA,\mu_B)} &=& \frac{1}{2} \left(
\langle \Phi | \beta_{\mu_B}  c_j c_i \beta_{\mu_B}^{\dagger}| \Phi \rangle
+
\langle \Phi | \beta_{\overline{\mu}_B}  c_j c_i \beta_{\overline{\mu}_B}^{\dagger}| \Phi \rangle
\right)
\nonumber\\
&=&
\left( V^* U^T \right)_{ij} +
\frac{1}{2}  \left( U_{i\mu_B}V^*_{j\mu_B}-V^*_{i\mu_B}U_{j\mu_B} \right)
\nonumber\\
&+&
\frac{1}{2}  \left( U_{i \overline{\mu}_B}V^*_{j \overline{\mu}_B}
-V^*_{i \overline{\mu}_B}U_{j \overline{\mu}_B} \right)
\end{eqnarray}
including not only the $\mu_B$ state but also its time reversed partner 
$\overline{\mu}_B$. The corresponding HF and pairing 
fields read
\begin{eqnarray}
\Gamma_{ij}^{(EFA,\mu_B)} &=& \sum_{qp} \overline{v}_{iqjp} \rho_{pq}^{(EFA,\mu_B)}
\end{eqnarray}

\begin{eqnarray}
\Delta_{ij}^{(EFA,\mu_B)} &=& \frac{1}{2} \sum_{qp} \overline{v}_{ijqp} \kappa_{pq}^{(EFA,\mu_B)}
\end{eqnarray}
and the total EFA energy can be written in the usual HFB form
\begin{eqnarray}
E^{(EFA,\mu_B)} &=& Tr \Big[ t \rho^{(EFA,\mu_B)}\Big] 
\nonumber\\
&+&
\frac{1}{2}  Tr \Big[ \Gamma^{(EFA,\mu_B)} \rho^{(EFA,\mu_B)}\Big] 
\nonumber\\
&-&
\frac{1}{2}  Tr \Big[ \Delta^{(EFA,\mu_B)} \kappa^{(EFA,\mu_B) *}\Big]
\end{eqnarray}

The microscopic justification of this expression for $E^{(EFA,\mu_B)}$
has been first given in Ref. \cite{perez} using ideas of 
quantum statistical mechanics. The EFA energy can be obtained as the
statistical average, with a given density matrix operator, and applying 
the variational principle to it, the usual HFB-EFA equation \cite{perez}
is obtained. The existence of a variational principle allows to
use the gradient method to solve the HFB-EFA equation \cite{perez} with the 
subsequent simplification in the treatment of the constraints.

\begin{figure}[ht]
\centering
\includegraphics[width=80mm]{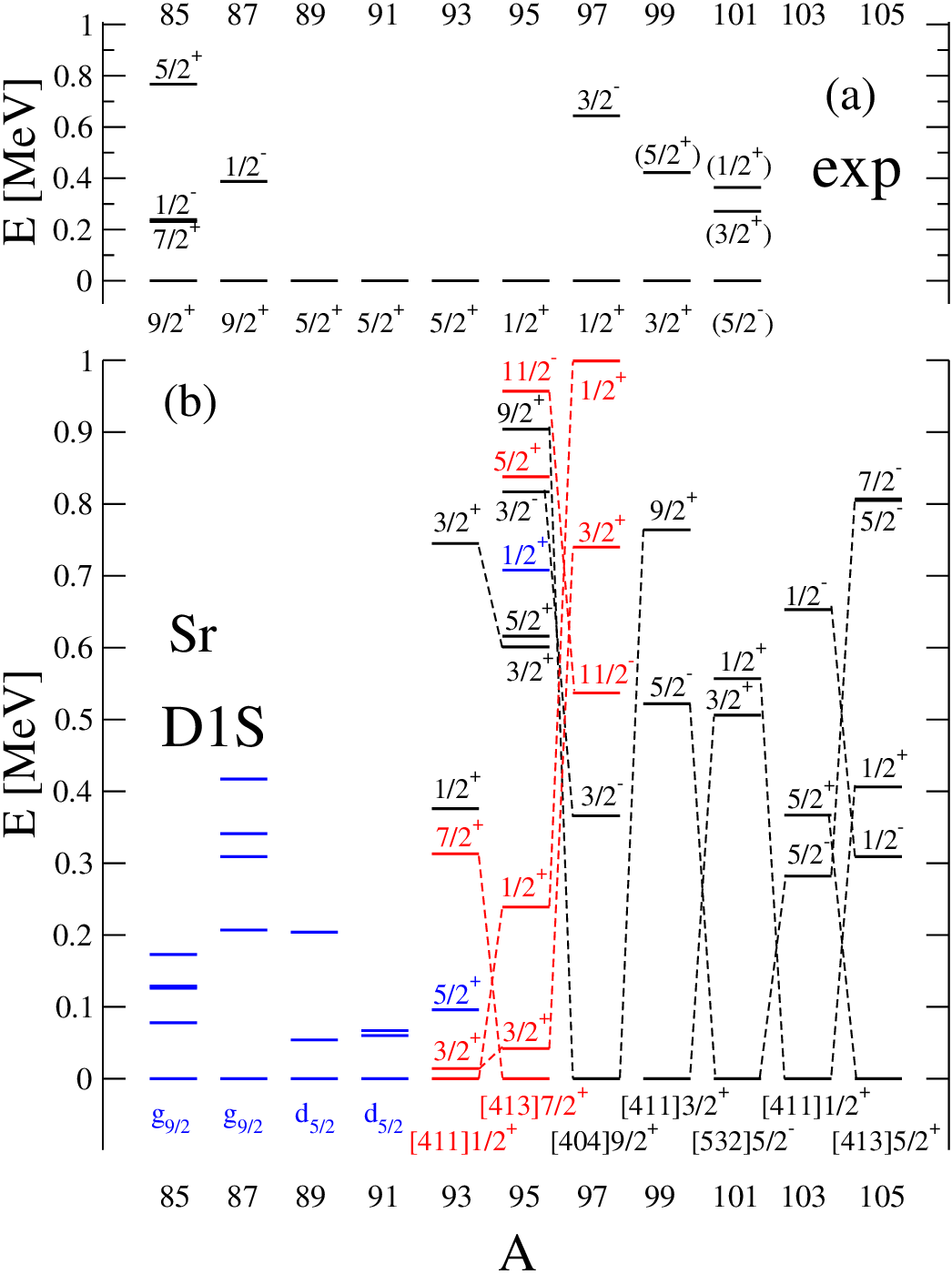}
\caption{(Color online) Experimental (a) excitation energies and spin-parity 
assignments of the non-collective states are compared
with  HFB-EFA results (b) for the  one-quasineutron states
 in odd-$N$ Sr isotopes (see text for details). Prolate 
 configurations are shown by 
 black lines, oblate ones by red lines, 
and spherical ones by blue lines.
}
\label{fig_sr}
\end{figure}

The solution of the HFB-EFA (as well as the exact blocked HFB) equation
depends upon the initial blocked level $\mu_B$. In the HFB-EFA case,
the $K$ quantum number is selfconsistently preserved along the calculation
and so does the parity if octupole correlations are not allowed  in the iterative process. Blocking levels with different 
$K$ values and parities lead to different quantum states of the odd-$A$
nucleus being the ground state the one with the lowest energy. Given
a $K$ value and parity, one could naively
think that considering the quasiparticle with the lowest 
quasiparticle energy as the initial blocked level should lead to the
lowest energy solution for that value of $K$ and parity. 
However, due to the selfconsistent nature of
the whole process this is not by any means guaranteed and therefore
several quasiparticles, usually corresponding to the lowest quasiparticle
energies have to be considered. In addition, in the present case and 
owing to the presence of coexisting prolate, oblate and spherical minima
in some of the nuclei considered, blocked configurations with those
quadrupole deformations have to be explored. For this reason
the minimization process has to be carried out 
several times  using different initial prolate, spherical and oblate 
(false) vacua. We have repeated each calculation, for a given 
false vacuum and   $K$-values from 1/2 up to 15/2, 
using as initial blocking states 
the 12 quasiparticles corresponding to the lowest quasiparticle energies.
The use of so many initial configurations is to  make sure we are not going to
miss the true ground state and all the lowest excited states. The resulting computational
cost is high as in the worst case (three false vacua) we have to carry 
out for each nucleus a total of $8\times 3 \times 12 = 288$ HFB-EFA calculations.
Among all the resulting one-quasineutron configurations, we have selected 
those within a 1 MeV energy window and compare them with available 
experimental data (see, for example, Figs. \ref{fig_sr} , \ref{fig_zr}, 
and \ref{fig_mo} below). Note  that for nuclei in this region of the 
nuclear chart, there exist several competing shapes at low excitation energy 
and therefore our procedure assures that the lowest energy solution can 
be reached for all values of the quadrupole moment $Q_{20}$ and mass number.
Finally, let us also mention that in those cases where the neutron 
pairing energy of a given false vacuum is too small, we increase it by 
constraining the mean value of the number particle fluctuation neutron 
operator $\left( \Delta N\right)^{2}$ in this false vacuum.
We have found that this procedure leads to a very fast convergence of the  calculations. 
  
%
%

\section{Results}
\label{RESULTS}


\subsection{Potential energy curves and single-particle energies}


It has been shown experimentally that the ground states of even-even 
Sr, Zr and Mo isotopes with $N$ ranging from the magic neutron  number $N=50$ 
up to $N\sim 60$ are weakly deformed, but they undergo a shape transition 
from nearly spherical to well deformed prolate (oblate) configurations 
as $N= 60$ is approached and crossed. Evidence for such an abrupt shape/phase 
transition includes $2^+$ lifetime measurements \cite{mach,goodin} and 
quadrupole moments for rotational bands \cite{urban}, as well as isotopic 
shifts in nuclear charge radii \cite{buchinger,campbell,charlwood}.
Heavier Sr and Zr ($A\sim 110$) isotopes exhibit  axially symmetric 
well deformed shapes, whereas the heavier Mo isotopes display a  tendency
towards triaxiality \cite{oursplb,hua}. Above this region, it has
 been suggested \cite{doba96} that 
the $N=82$ shell closure might be quenched far from stability 
but still weak experimental evidence has been found.

In Fig. \ref{fig_pec}, we show the PECs for the even-even Sr (a), Zr (b), 
and Mo (c) isotopes. We display PECs corresponding to 
$N=$52, 56, 60, 64 and 68 isotones from which,  the main features 
of the shape evolution can be followed. Nuclei with $N=52$ show 
a sharp  spherical minimum 
that becomes rather shallow at $N=56$. Isotopes with $N=60$ are already 
deformed with oblate and prolate minima very close in energy. In the 
case of Sr isotopes the ground state is prolate, for Zr isotopes  
oblate and prolate minima are almost degenerate, while 
the ground state becomes  oblate for Mo nuclei. For neutron numbers
greater than 60, prolate and oblate minima become well defined. 
Our results agree well
with the shape/phase change predicted 
around $N = 60$ by microscopic-macroscopic models 
\cite{skalski,xu02,moller95,moller08} and microscopic selfconsistent 
relativistic \cite{lala} and non relativistic \cite{bender06,delaroche} 
mean field calculations. Later on, in section 
\ref{one-qp-Sr-Zr-Mo}, our analysis of the shape evolution in
even-even Sr, Zr and Mo nuclei will be further extended with the 
systematics of the ground and low-lying one-quasineutron 
configurations predicted by our HFB-EFA
calculations for  odd-$A$ nuclei in these isotopic chains. 

\begin{figure}[ht]
\centering
\includegraphics[width=80mm]{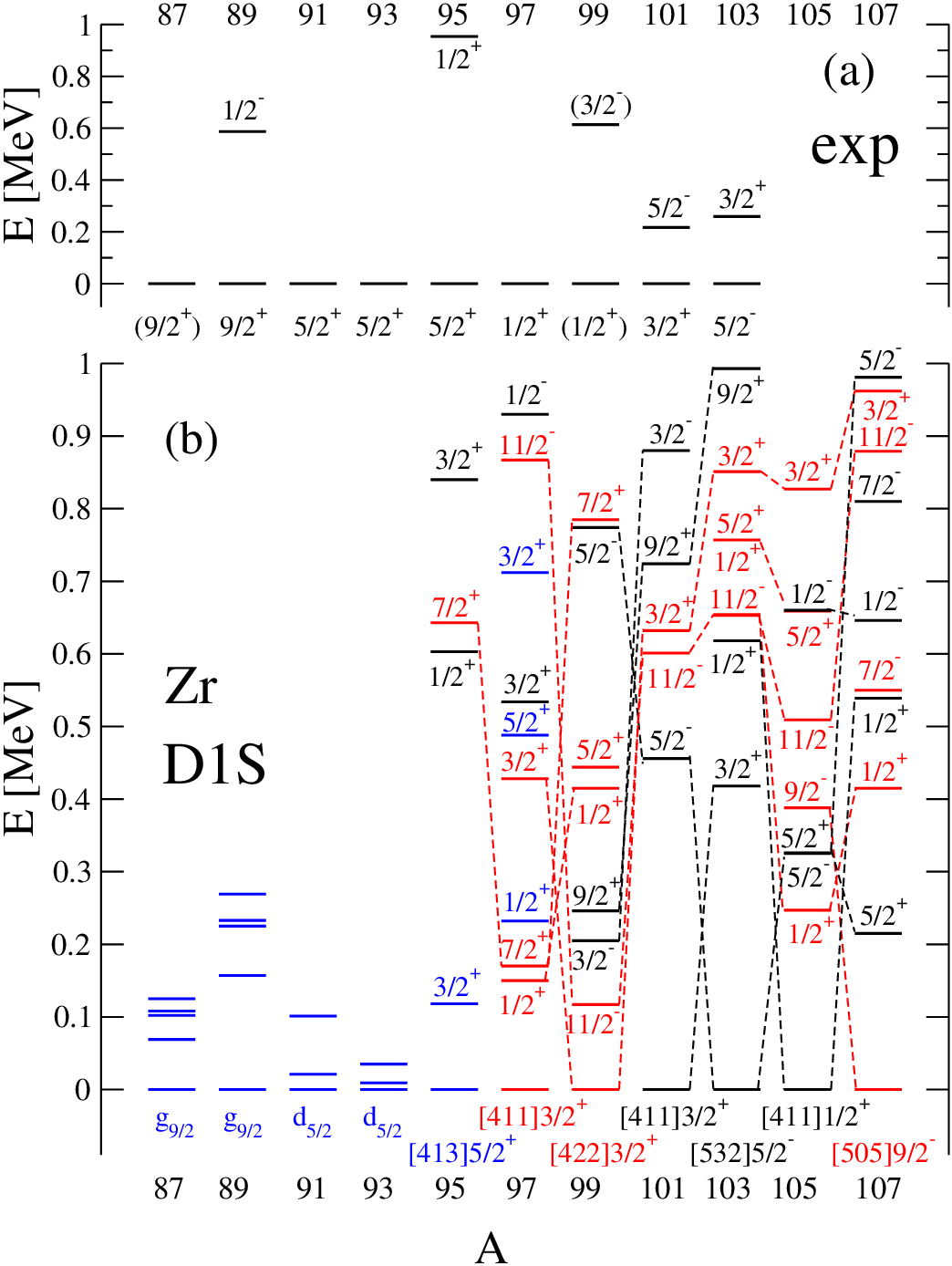}
\caption{(Color online) Same as in Fig. \ref{fig_sr}, but for odd-$N$ Zr isotopes.}
\label{fig_zr}
\end{figure}

Another interesting outcome of our HFB calculations for even-even
nuclei (see, also the discussion in section 
\ref{one-qp-Sr-Zr-Mo}), are the proton and neutron
single-particle energies (SPEs) shown in Fig. \ref{fig_spe}, as
 functions of the axial quadrupole 
moment $Q_{20}$, for the case of $^{102}$Zr ($Z=$40, $N=$62). To obtain 
these Nilsson-like diagrams, we have computed  the eigenvalues
of the Routhian $h = t+ \Gamma - {\lambda}_{20}Q_{20}$, with $t$ being 
the kinetic energy operator, $\Gamma$ the Hartree-Fock field and 
${\lambda}_{20}Q_{20} $ the term containing the Lagrange multiplier 
used to enforce the corresponding quadrupole constraint. Obviously, the 
usual mean field constraints \cite{rs} on both neutron and proton numbers 
are also taken into account. Fermi levels are plotted with 
thick dashed (red) lines
in the figure. These type of diagrams help us \cite{PT-D1M,ours2}
to identify  regions of low level density around the corresponding Fermi levels 
which, according to the 
Jahn-Teller effect \cite{JTE-1}, favor the onset of deformation.

For axially symmetric configurations, the SPE levels are tagged by the 
(half integer) K quantum number that corresponds to the third component
of the angular momentum in the intrinsic frame. Due to the time-reversal
invariance, single-particle orbitals with the same absolute K value
are degenerate (Kramers degeneracy). Positive (negative) parity levels 
are plotted with solid (dashed) lines. At the spherical 
configuration $Q_{20}=0$, the quantum numbers $nlj$ are recovered. 
The asymptotic quantum numbers $[N,n_z,\Lambda]K^\pi$ are shown
in Fig. \ref{fig_spe} for  $Q_{20}$ values close to those where the 
minima of the PECs (see, Fig. \ref{fig_pec})
are located (vertical arrows). For each single particle state they have been
assigned according to the largest amplitude in the expansion of the single
particle wave function in the axially symmetric harmonic oscillator basis.

For the neutron-rich $A\sim 100$ nuclei of our concern in the present work, the 
valence protons occupy the $N=3$ shell and start to fill the 
$g_{9/2}$ orbitals in the $N=4$ shell. Neutrons occupy states belonging to the $N=4$ shell, 
which are strongly mixed by deformation, and begin to fill the $h_{11/2}$ 
intruder orbitals coming down from the $N=5$ oscillator shell. Because 
of their opposite parity, the $h_{11/2}$ neutron states (the $g_{9/2}$ 
proton states) do not mix with the $N=4$ neutron orbitals (the $N=3$ 
proton orbitals). These intruder orbitals polarize the core towards 
oblate and prolate deformations. As a consequence, the underlying nuclear 
structure in this mass region is very sensitive to the occupancy of these 
single-particle orbitals and the result is a rapid change in nuclear 
spectroscopic properties as a function of both neutron and proton 
numbers. 

Below we compare the results obtained with the D1S parametrization of the
force with those obtained with the newly introduced D1M parametrization. 
As it turns out that the spin orbit strength of D1M is 12 $\%$ smaller 
than the one of D1S
one may wonder whether this difference could lead to a strong impact in the
single particle spectrum and its behavior with deformation. To answer this
question we have plotted the D1M single particle levels and compared them
to the D1S ones. The comparison of the two sets of levels shows  small
differences in the position of the levels (of the order of a few hundred
of keV at most) that are not strong enough as to change the ordering of
the spherical single particle spectrum. The behavior with deformation
is not changed in a significant way either. This small impact can be 
related to the larger effective mass of D1M as compared to D1S, 
that makes the single particle spectrum more dense in D1M than in D1S.
As a consequence a weaker spin-orbit strength is required in D1M to shift 
the intruders to the lower shell.


\subsection{One-quasiparticle states in odd-$N$ Sr, Zr and Mo nuclei}
\label{one-qp-Sr-Zr-Mo}

\begin{figure}[ht]
\centering
\includegraphics[width=80mm]{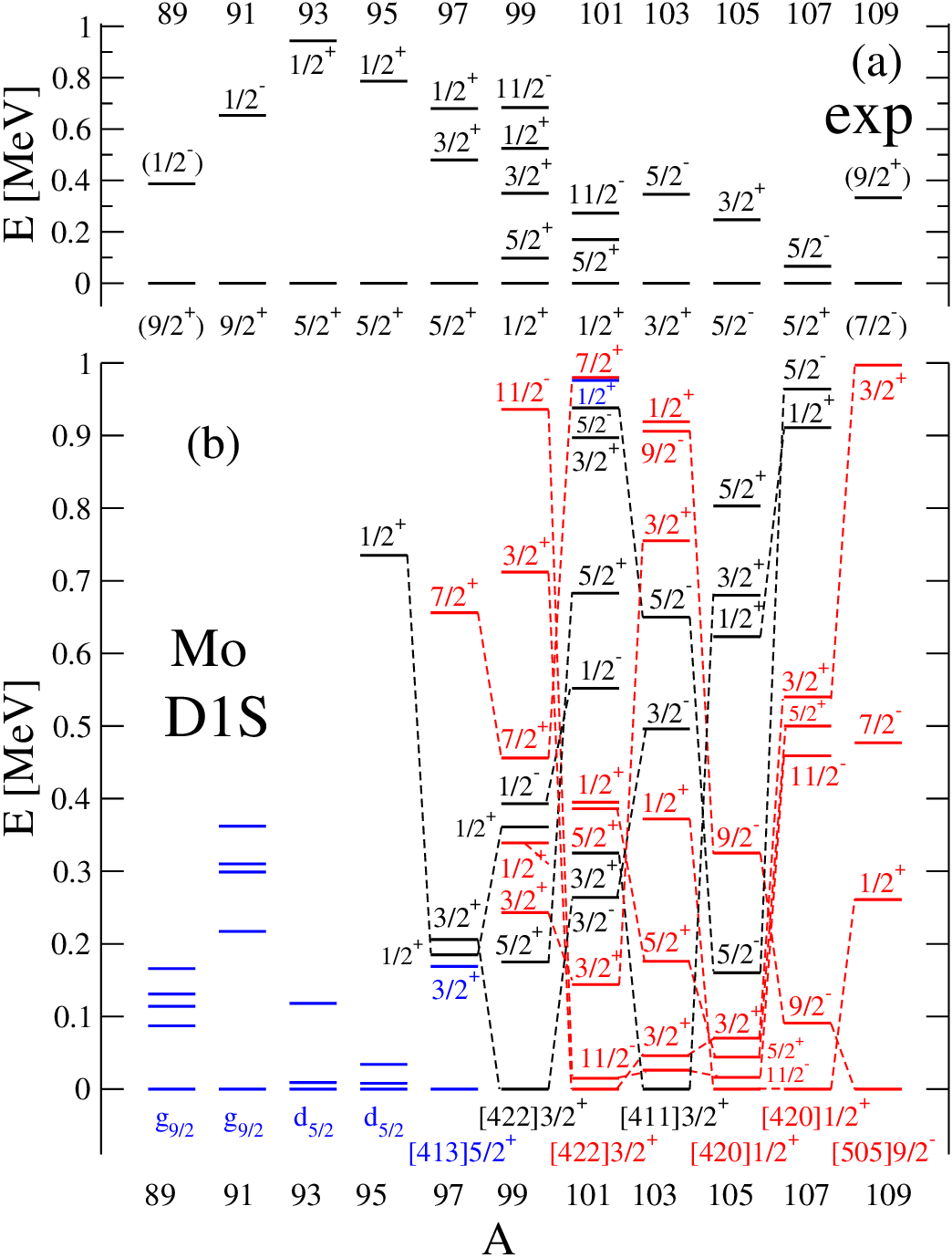}
\caption{(Color online) Same as in Fig. \ref{fig_sr}, but for  odd-$N$ Mo isotopes.}
\label{fig_mo}
\end{figure}

In Figs. \ref{fig_sr}, \ref{fig_zr} and \ref{fig_mo} results are given for 
odd-$N$ Sr, Zr, and Mo isotopes, respectively. In each figure the experimental 
\cite{exp_spect} ground and low-lying one-quasineutron states, characterized by their 
spin-parity assignments (panels (a) of the figures) are compared  with 
the ones predicted within our HFB-EFA calculations (panels (b) of the figures). 
This is certainly a very challenging test for the spectroscopic quality of 
the Gogny-D1S EDF for which results are shown. It is worth noting that 
in some cases the experimental spin and parity assignments in panels (a) 
have been deduced from systematics and should be treated with caution.

The existence of oblate and prolate competing shapes in this mass region 
prevents us to use the usual convention \cite{naza90,nielsen} of plotting 
the theoretical results with hole (particle) states below (above) the 
corresponding
ground state. In Figs. \ref{fig_sr}, \ref{fig_zr} and \ref{fig_mo}, the excited states in a given isotope 
are referred to the corresponding  ground state  regardless its shape.
Prolate configurations are shown by black lines, oblate ones by red lines, 
and spherical ones by blue lines. The quasiparticle states in the deformed 
configurations are labeled by their $K^\pi$ quantum numbers 
and, in addition, the ground states are labeled by their asymptotic (Nilsson) quantum numbers 
$[N,n_z,\Lambda]K^\pi$. In the lighter isotopes we obtain 
spherical (sharp or shallow) equilibrium shapes and therefore only  the spherical 
orbital, which is either $g_{9/2}$ or $d_{5/2}$ in all the cases, is indicated.
Like-orbitals in 
neighbor isotopes are connected by dashed lines to better appreciate  their 
isotopic evolution.

Looking at Figs.  \ref{fig_sr}, \ref{fig_zr} and \ref{fig_mo}, 
the striking  correlation between the ground state deformations  
in the even-even isotopes (Fig. \ref{fig_pec}) and the ground
state shapes of the odd ones is observed. 
The lighter Sr nuclei ($A=85-91$) are spherical. They  become oblate 
for $A=93$ and 95 with 
prolate states at low excitation energies. Finally, prolate 
deformed ground states are obtained 
in the heavier isotopes with $A=97-105$.  Oblate excited states 
below 1 MeV are obtained for $A=97$, whereas they lie above 1 MeV
in the heavier ones.

The lighter Zr isotopes ($A=87-95$) displayed in Fig. \ref{fig_zr}, 
show  spherical equilibrium shapes, which turn into oblate ($A=97,99$), 
prolate ($A=101-105$), and again oblate ($A=107$) ground states with 
excited one-quasiparticle configurations of different shapes below 
1 MeV and very close in energy, in agreement with Fig. \ref{fig_pec} (b).

The odd Mo isotopes in Fig. \ref{fig_mo} also display similarities 
with Fig. \ref{fig_pec} (c). Again, spherical ground states are found for  
the lighter isotopes. Alternating oblate and prolate equilibrium  shapes
are predicted in  
$^{99,101,103,105}$~Mo. Prolate and oblate configurations   
are particularly close in energy
within this mass range. Oblate ground state deformations are found for both 
$^{107,109}$~Mo with 
no prolate one-quasineutron 
configurations  in $^{109}$~Mo below 1 MeV.

In Fig. \ref{fig_sr}, we compare the experimental (a) low-lying one-quasineutron
spectra with the ones predicted by our HFB-EFA calculations (b)
for  odd-$N$ Sr isotopes. We can see a reasonable 
agreement in the ground states, except for the isotopes $^{93,95,97}$Sr.
In  $^{93}$Sr, the $5/2^+$ experimental ground state appears at 0.1 MeV
in our calculations, whereas in $^{95,97}$Sr the $1/2^+$ experimental ground
states are found as oblate excited configurations at $E=0.25$ and $E=1$ MeV,
respectively. The ground states of the heavier isotopes 
$^{99,101}$Sr are well reproduced.

The structural  evolution of the 
ground and excited one-quasiparticle states 
along the considered Sr, Zr and Mo isotopic chains
can also be followed 
by looking at the dashed lines connecting the 
states characterized by the same asymptotic 
quantum numbers $[N,n_z,\Lambda]K^\pi$ 
and  at 
Fig. \ref{fig_spe}, as a guide. Indeed, the  
ground state sequence as the number of neutron increases
exhibits a clear correspondence with the orbitals found in Fig. \ref{fig_spe}
along thick vertical arrows at 
oblate ($Q_{20}\sim -2.5$ b)  and prolate  ($Q_{20}\sim 5$ b) 
quadrupole moments. For example, one sees  how the ground state in 
$^{101}$Sr, which corresponds
to the prolate $[532]5/2^-$ state, becomes 
the first 
one-quasineutron 
excitation, below 0.6 MeV, in 
the neighboring $^{99}$Sr and $^{103}$Sr with 
prolate $[411]3/2^+$ and $[411]1/2^+$ ground states. 
On the other hand, 
$^{97}$Sr exhibits a prolate $[404]9/2^+$ ground state, which turns
 out to be a quasineutron band head in  $^{95}$Sr and 
$^{99}$Sr while the $[411]1/2^+$ prolate ground state in $^{103}$Sr
appears  as an excited 
state between 0.6 and 0.4 MeV in $^{101}$Sr and  $^{105}$Sr, respectively

In Fig. \ref{fig_zr} the comparison is made for Zr isotopes. The
 agreement
is again very reasonable exception made of $^{97,99}$Zr 
for which the 
reported ground states appear in our HFB-EFA calculations as oblate 
excited states below 0.5 MeV. Special mention 
deserves, the remarkable agreement for the 
heavier isotopes $^{101,103}$Zr, where our calculations reproduce 
both
the 
ground states and the low-lying excitations. One can 
see, how the band 
head $5/2^-$ at 0.45 MeV in $^{101}$Zr and the ground state  in 
$^{103}$Zr
correspond to the same $[532]5/2^-$  single-particle configuration,
as it has been experimentally established in Ref. \cite{hua}.

The dashed lines in Fig.\ref{fig_zr} are also useful to trace 
the isotopic evolution of the 
single-particle configurations, which can be also understood by
moving the (neutron) Fermi level in Fig. \ref{fig_spe} through 
the vertical arrows.
Thus, taking for example the prolate solutions, if one looks at the 
orbitals intersecting the vertical arrow at 
$Q_{20}=5$ b in Fig. \ref{fig_spe}, one finds the deepest state at 
about $e_{sp}=-9$ MeV, which corresponds 
to a $[420]1/2^+$ $(s_{1/2})$ 
state. It can be seen, in Fig.\ref{fig_zr}, as the 
prolate excited state at 0.6 MeV in $^{95}$Zr.
The next prolate states in Fig. \ref{fig_spe} are $[550]1/2^-$ falling 
down very quickly from $h_{11/2}$ and $[422]3/2^+$ $(d_{5/2})$
appearing as the first prolate excitations in $^{97}$Zr.
On the other hand, in $^{99}$Zr the first two prolate states
$3/2^-$ and $9/2^+$ appear around 
0.2 MeV. They are  
associated with the  $[541]3/2^-$ coming down from $h_{11/2}$ 
and $[404]9/2^+$ raising from   $g_{9/2}$ in Fig.  \ref{fig_spe}. 
Going further up in 
Fig. \ref{fig_spe} we find $[411]3/2^+$ $(g_{7/2})$, which is the ground 
state in $^{101}$Zr and  $[532]5/2^-$ $(h_{11/2})$ which 
represents an excited 
state at 
about 0.45 MeV. Further up in excitation energy one finds, the  $9/2^+$
and $3/2^-$ states already discussed (see the connections with dashed lines
with their partners in  $^{99}$Zr). 
In  $^{103}$Zr the $3/2^+$ and
$5/2^-$ states interchange their positions with respect
to  $^{101}$Zr and, as it can be easily understood from
Fig. \ref{fig_spe}, the $[532]5/2^-$ configuration becomes now
the ground state. Subsequently, the 
$[411]1/2^+$ $(g_{7/2})$ ground state 
is predicted by our HFB-EFA. The excited one-quasineutron 
$5/2^-$ and $5/2^+$ states,  at about 0.35 MeV, correspond
to the 
already discussed  $[532]5/2^-$ $(h_{11/2})$, which now lies below the
Fermi level, and to the   $[413]5/2^+$ $(d_{5/2})$ configuration, above the
Fermi level, that becomes the lowest prolate deformed
one-quasiparticle excitation in the last  odd-$N$ nucleus
considered (i.e., $^{107}$Zr). 

A similar analysis can be made for the oblate 
one-quasineutron
states by looking at the 
intersections of the single-particle levels in Fig.  \ref{fig_spe}
with the vertical line at $Q_{20}=-2.5$ b. As can be seen from 
Fig. \ref{fig_zr}, $^{97}$Zr displays a ground 
$3/2^+$ and two low-lying excited ($1/2^+$ 
and $7/2^+$) oblate deformed states. They 
can be  associated with the three orbitals 
at -8 MeV in the oblate sector of  Fig.  \ref{fig_spe}, namely,
 $[411]3/2^+$ $(d_{5/2})$,  $[411]1/2^+$ $(d_{5/2})$, and  
$[413]7/2^+$ $(g_{7/2})$. Next, in $^{99}$Zr we find
an oblate $3/2^+$ ground state and the oblate $11/2^-$ excitation
at 0.1 MeV which correspond  to the $[422]3/2^+$ $(g_{7/2})$
and $[505]11/2^-$ intruder state coming down from the  $h_{11/2}$
shell. The  $[420]1/2^+$  and 
$[422]5/2^+$ (coming from $g_{7/2}$ shell) excitations , are also  
visible in the quasiparticle spectrum of $^{99}$Zr. 
The nucleus $^{101}$Zr exhibits a prolate ground state but one finds
a  $[505]11/2^-$ as the lowest oblate quasineutron configuration 
very close
to the $3/2^+$ excitation (that was the ground state in  $^{99}$Zr).
The same comments apply to both   $^{103}$Zr and  $^{105}$Zr
but now the state  $[420]1/2^+$ $(s_{1/2})$ comes into play.
Finally, $^{107}$Zr displays an oblate ground state that 
can be associated with a $[505]9/2^-$  configuration 
coming down from the $h_{11/2}$ shell.
  
\begin{figure}[ht]
\centering
\includegraphics[width=80mm]{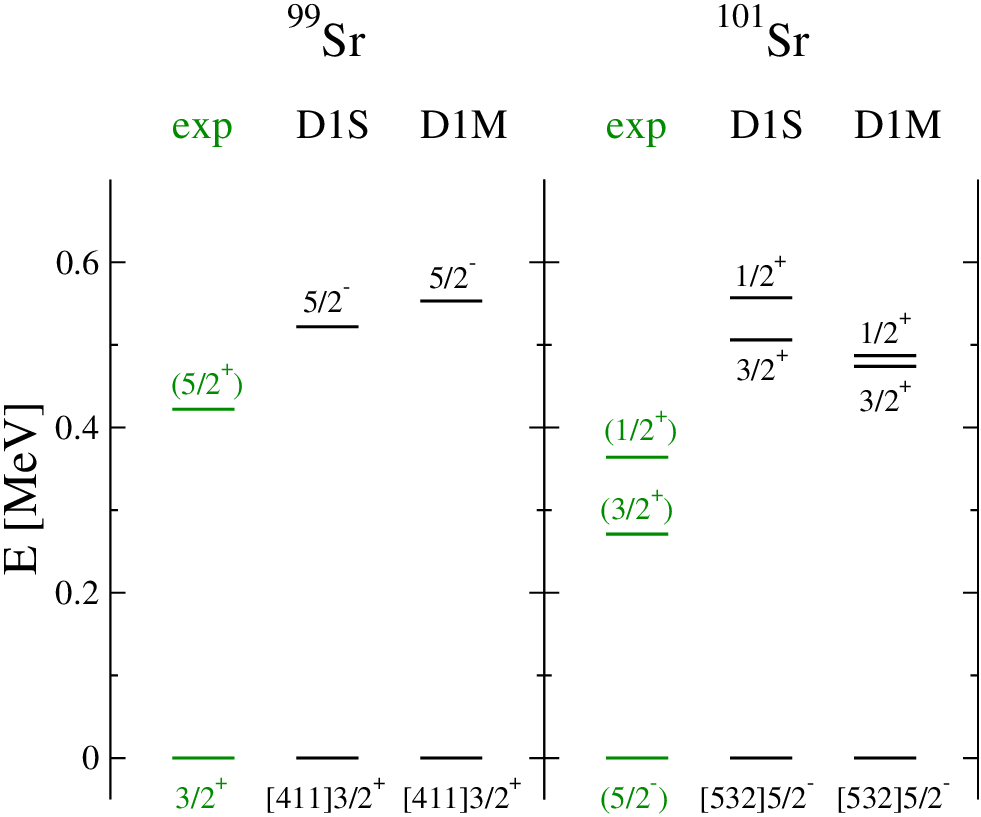}
\caption{(Color online) Excitation spectra of some Sr isotopes obtained 
within the HFB-EFA calculations based on the D1S and
D1M parametrizations of the Gogny-EDF are compared with available experimental
data. The color scheme used to characterize the quadrupole deformation of
each of the states is the same as in Fig. \ref{fig_sr}. 
Experimental levels are plotted in green because there is no clear
experimental evidence concerning the sign of their intrinsic deformations.}
\label{sr_comp}
\end{figure}

According to the discussions in Ref. \cite{hua}, where the rotational bands
of the lowest quasiparticle states were measured, and the results 
of our recent study in Ref. \cite{oursplb}, where 
triaxial shapes were predicted in neutron-rich
Mo isotopes, the odd Mo nuclei are the most difficult to analyze. 
Our HFB-EFA description in terms of axial shapes nicely reproduces 
the ground $5/2^+$ and the two
excited $3/2^+$ and $1/2^+$ states in $^{97}$Mo. Our calculations 
also provide us with the observed  $1/2^+$, $5/2^+$, $3/2^+$, and  $11/2^-$ 
one-quasiparticle configurations 
in $^{99}$Mo although redistributed. The observed
$1/2^+$, $5/2^+$, and  $11/2^-$ states are obtained 
in $^{101}$Mo
within 0.4 MeV excitation energy. On the other hand, the
ground state in $^{103}$Mo
is well reproduced while the excited $5/2^-$ appears at about 0.6 MeV.
The latter becomes the experimental ground state in $^{105}$Mo 
but it is predicted at 0.15 MeV in our calculations.
Finally, the observed ground states in both 
$^{107}$Mo and $^{109}$Mo, are obtained as oblate
excited (about 0.5 MeV) configurations within our 
HFB-EFA calculations.

In general, we observe that the main difficulties to describe the
experimental information appear in those cases where the mean field
approach might not  be sufficient. Such is the case of $N=55-59$ isotopes,
where we get very shallow minima (see Fig. \ref{fig_pec}), 
a situation that in general requires configuration 
mixing for a better description. Work along this direction has already
been carried out in \cite{Vermeulen.07,Gaudefroy.09} in the context of
a collective hamiltonian obtained in the Gaussian Overlap Approximation
(GOA) framework. We believe that a better treatment of the problem is
required to pin down the subtle details of the spectrum of an odd-A system
and therefore the exact Generator Coordinate Method (GCM) with the axial quadrupole
moment would be required. Work along this line is in progress and will be
reported elsewhere. Difficulties also appear  when the description requires
the consideration of triaxility as it is the case for the heavier Mo isotopes, 
which exhibit a 
$\gamma$-soft behavior and even triaxial minima \cite{oursplb} with the 
axial ones converted into saddle points. In such a situation one also 
expects that the oblate and prolate solutions will be highly mixed
with the intermediate triaxial configurations.
On the other hand, a HFB axial description is expected to work better for 
well developed axial minima separated by energy barriers in both spherical 
and triaxial shapes as in the heavier isotopes of Sr and Zr.

\begin{figure}[ht]
\centering
\includegraphics[width=80mm]{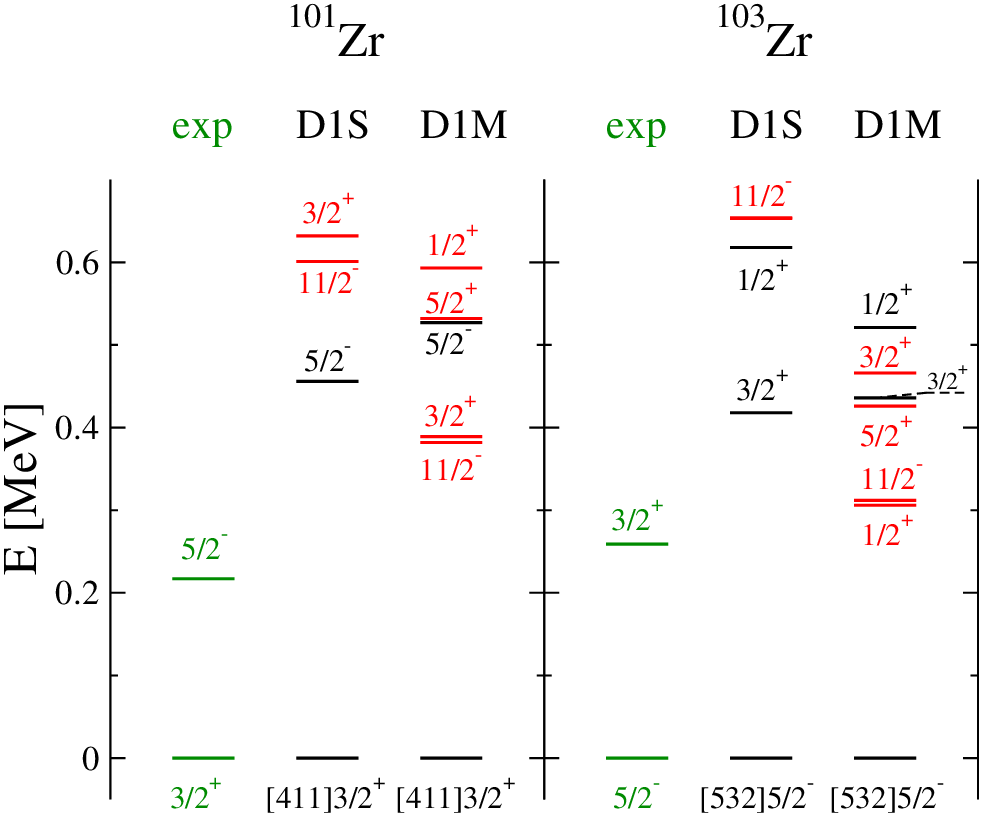}
\caption{(Color online) Same as in Fig. \ref{sr_comp}, but for Zr isotopes.}
\label{zr_comp}
\end{figure}

Another interesting point worth noticing is the
isotonic behavior of the considered odd-$A$ 
nuclei. Because the isotopes studied in 
this work have an odd number of neutrons, one observes very similar 
spectroscopic schemes for fixed number of neutrons (isotones), where the 
odd neutron determines the spin and parity of the ground state.
First, one realizes the observed experimental
correspondence between the spins and parities of isotones within Sr, Zr, 
and Mo nuclei. Thus, starting at $N=47$ with $J^\pi = 9/2^+$ in 
$^{85}$Sr,  $^{87}$Zr, and $^{89}$Mo, the sequence of 
$9/2^+,9/2^+,5/2^+,5/2^+,5/2^+,1/2^+,1/2^+,3/2^+,5/2^-$ is
observed experimentally from $N=47$ up to $N=63$.
The agreement between these experimental findings and our calculations is 
specially remarkable in the heavier isotones.

For $N=61$ ($^{99}$Sr,  $^{101}$Zr, $^{103}$Mo) the ground state
corresponds to $[411]3/2^+$ states both experimentally and theoretically.
In addition, a low-lying band head $5/2^-$ is observed experimentally \cite{hua} 
in  $^{101}$Zr and $^{103}$Mo that corresponds reasonably well 
with the predictions of our HFB-EFA
calculations. The experimental assignment  $(5/2^+)$ to the excited 
state in $^{99}$Sr is uncertain and could also correspond to a $5/2^-$
in which case the similarity would be complete.
The $N=63$ isotones $^{101}$Sr and $^{103}$Zr exhibit
a $[532]5/2^-$ ground state in agreement with experiment. In the case
of $^{105}$Mo we obtain an oblate ground state $[420]1/2^+$, which does not
show up experimentally, but the first excited prolate state is again 
$[532]5/2^-$. Excited $3/2^+$ quasiparticle configurations
observed experimentally
at $N=63$ are also well reproduced in the calculations as prolate configurations.
The $N=65$ isotones  $^{103}$Sr and  $^{105}$Zr are predicted to be prolate
$[411]1/2^+$ in their ground states, while  $^{107}$Mo is predicted
as an oblate $[420]1/2^+$ at variance with experiment.

The main difficulty to understand the previous correspondence appears in the 
isotones $N=$57, 59, where the experimental $1/2^+$ assignments to their
ground states are difficult to reproduce in our calculations.
However, these $1/2^+$ quasiparticle states appear in the calculations as 
excited oblate configurations at low excitation energies.
The $3/2^-$ excited state observed experimentally in $^{97}$Sr and
$^{99}$Zr appear in our calculations at similar energies as prolate 
configurations. In $^{99-101}$Mo there are two excited 
$5/2^+$ and $11/2^-$ 
states also found  in our HFB-EFA calculations below the 1 MeV energy window 
of the plot.

\begin{figure}[ht]
\centering
\includegraphics[width=80mm]{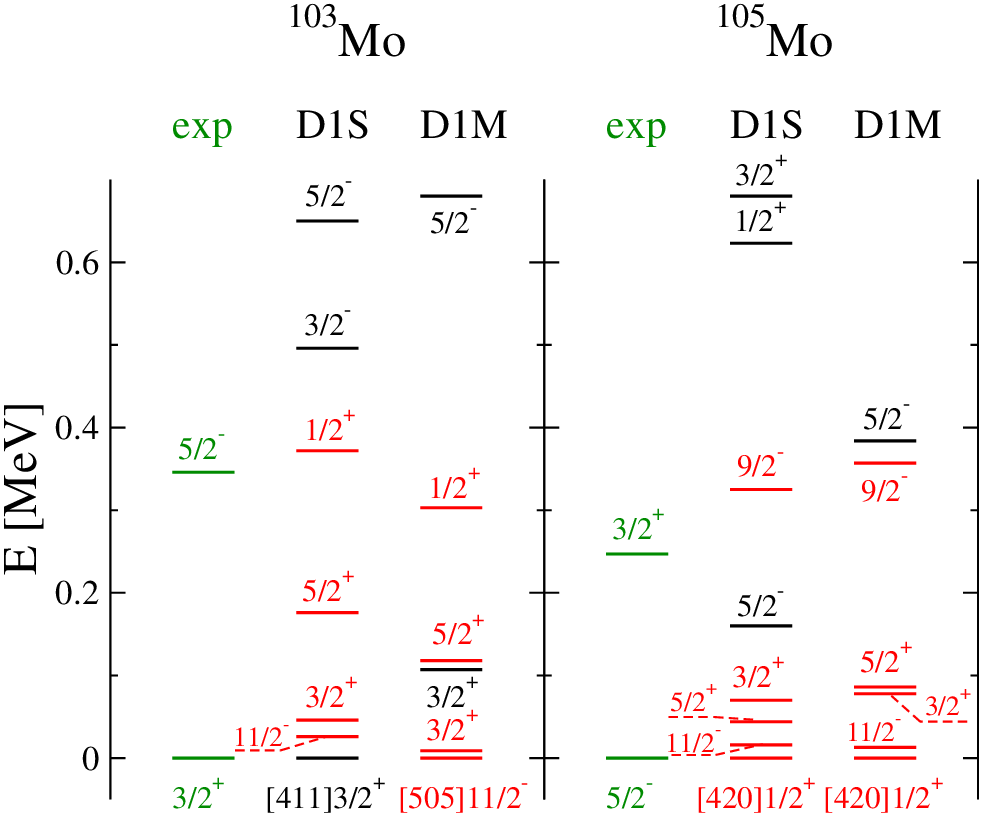}
\caption{(Color online) Same as in Fig. \ref{sr_comp}, but for Mo isotopes.}
\label{mo_comp}
\end{figure}

As mentioned in Sec. \ref{INTRO}, one of our aims 
in the present study was to further extend the 
comparison, already undertaken in Ref. \cite{PT-D1M}, between
the performance of the parametrizations Gogny-D1S and Gogny-D1M 
to the description of the spectroscopy of odd-$A$ nuclei. 
Therefore, in Fig. \ref{sr_comp} (for $^{99,101}$Sr),  Fig. \ref{zr_comp} (for 
$^{101,103}$Zr) and  Fig. \ref{mo_comp} (for $^{103,105}$Mo), we compare
the results obtained for the low-lying one-quasineutron
states using these two different incarnations of the 
Gogny-EDF with the experimental data. First, one realizes the nice similarity
between
the HFB-EFA predictions obtained with these two parametrizations. In 
particular, the predicted ground states coincide in most of the cases
and agree well with experiment. Only in Mo isotopes
one observes some 
disagreement  but in this case, we obtain a high density of very low-lying
configurations with the observed ground states always below 0.2 MeV. 

Looking at the figures in more detail, we can see
that within the considered  energy window  of 0.7 MeV, practically the same states
are obtained from the calculations with D1S and D1M. This is
especially true for prolate deformed states. For example, in Sr 
isotopes (Fig. \ref{sr_comp}) we obtain the same  $5/2^-$ states
in  $^{99}$Sr as well as the   $3/2^+$ and   $1/2^+$ states in  $^{101}$Sr
very close to the experiment and with very small variations (less than
0.2 MeV) from one EDF to another. 

The same happens in Fig. \ref{zr_comp}
for the Zr isotopes. The  prolate states $5/2^-$ (in $^{101}$Zr)
and 
$3/2^+$,  $1/2^+$ (in $^{103}$Zr) are predicted  very close with D1S and D1M 
and agree well with experiment.
 However, we observe an enhanced sensitivity
in the case of the oblate configurations, resulting in lower energies, of 
the order of a few hundred keV, in the case of D1M with respect 
to D1S. This is a consequence of having closer oblate and prolate 
minima with the parametrization D1M. The ground states are nicely reproduced.

In the case of Mo isotopes (Fig. \ref{mo_comp}) the
experimental ground  and first excited state are interchanged in $^{103}$Mo 
and $^{105}$Mo. Our HFB-EFA calculations provide prolate $3/2^+$ and $5/2^-$
states compatible with experiment, but also other oblate configurations 
at very low excitation energy. We can see  $11/2^-$, $3/2^+$,
 $5/2^+$, and $1/2^+$ low-lying oblate states in $^{103}$Mo. In addition
to those we also find an oblate $9/2^-$ state in  $^{105}$Mo.  As for
Zr isotopes, we also find that D1M tends to favor oblate solutions, which appear
displaced to lower energies with respect to the prolate solutions in D1S.
In general we obtain a comparable spectroscopic quality of the two Gogny 
parametrizations and it is  very satisfying  to observe how the
new parametrization D1M, in spite
of the relaxation of some of the original 
constraints in its fitting protocol and more 
oriented to reproducing nuclear
masses \cite{d1m}, still follows 
very closely the fine details predicted 
with Gogny-D1S for odd-$A$ nuclei in a region of the nuclear chart 
with such a challenging shape evolution.

\section{Conclusions}
\label{CONCLU}

In this work we have studied the
systematics of one-quasineutron configurations in 
odd-$A$ Sr, Zr and Mo isotopes within the selfconsistent 
HFB plus EFA framework, an approach that has been shown to 
be fairly adequate for most purposes. However, we are aware of the 
challenge of reproducing in detail the observed spectroscopic properties in the
particular mass region considered in the present study. Therefore, 
our aim has been to understand qualitatively the structural 
evolution of the ground and low-lying one quasiparticle 
configurations with neutron number. We have also shown that the
quality of the spectroscopic results obtained with the recent 
D1M parametrization of the Gogny force is comparable to the one
obtained with the standard parametrization D1S. We conclude that
both D1M and D1S parametrizations reproduce, at least qualitatively,  
the main features observed in the isotopic (and isotonic) trends of the
neutron-rich and odd-$A$ Sr, Zr and Mo nuclei. 

The main deficiency of our mean field description is the preservation
of axial symmetry. We have found this to be relevant in nuclei  
characterized by shallow minima around the spherical configurations 
like in the  $N$=55-59 isotopes,
as well as in those  nuclei with some tendency to $\gamma$-soft 
behavior or even triaxial minima like in the heavier Mo isotopes \cite{oursplb}. 
A beyond mean-field treatment seems to be necessary to improve the quality of the 
description in those cases, and configuration mixing calculations 
in the spirit of the Generator-Coordinate-Method (GCM) method  
\cite{rs} may be  required to improve the mean-field results. 
We are still far from being able to apply such a configuration mixing 
approach within an exact blocking scheme for odd nuclei. 
Nevertheless, a HFB-EFA-GCM scheme can be considered as a plausible 
step forward in this direction and work along this lines is in progress.
  

\noindent {\bf Acknowledgments}

This work was supported by MICINN (Spain) under research grants 
FIS2008--01301, FPA2009-08958, and FIS2009-07277, as well as by 
Consolider-Ingenio 2010 Programs CPAN CSD2007-00042 and MULTIDARK 
CSD2009-00064. The first steps of this work were
undertaken in the framework of the FIDIPRO program (Academy of 
Finland and University of Jyv\"askyl\"a) and one of us (R.R)
thanks Profs. J. Dobaczewski and  J. \"Aysto and the experimental 
teams of the University of Jyv\"askyl\"a (Finland) for 
warm hospitality and encouraging discussions.

\end{document}